\documentclass[]{mn2e}

\newif\ifAMStwofonts

\usepackage{graphicx}

\def\bib{\parskip=0pt\par\noindent\hangindent\parindent \parskip =2ex plus .5ex
    minus .1ex}

\newcommand{\gsim}{\raisebox{-0.13cm}{~\shortstack{$>$ \\[-0.07cm] $\sim$}}~}

\title{A deeper view of Extremely Red Galaxies: the redshift distribution in the
       GOODS/CDFS ISAAC field.}

\author[K.I. Caputi, J.S. Dunlop, R.J. McLure and N.D. Roche] {K.I. Caputi
          \thanks{kic@roe.ac.uk}, J.S. Dunlop \thanks{jsd@roe.ac.uk},
          R.J. McLure \thanks{rjm@roe.ac.uk} and N.D. Roche
          \thanks{ndr@roe.ac.uk}
       \\ Institute for Astronomy, University of Edinburgh, Royal Observatory,
       Edinburgh EH9 3HJ, Scotland, U.K. \\ }

\date{ }

\begin{document}

\maketitle

\label{firstpage}

\begin{abstract}
 We have analysed 5-epoch GOODS HST-ACS B, V, $\rm I_{775}$ and z 
 datasets (V1.0 release), in conjunction with existing VLT-ISAAC imaging in 
 the J, H and $\rm
 K_s$ bands, to derive estimated redshifts for the sample of 198 Extremely Red
 Galaxies (ERGs) with $\rm K_s<22$ (Vega) and $\rm (I_{775}-K_s)>3.92$ selected
 by Roche et al. (2003) from 50.4 arcmin$^2$ of the GOODS/CDFS field. We find
 that, at this depth, the ERG population spans the redshift range $\rm
 0.5<z_{phot}<4.75$ and over two decades in mass ($\rm \sim 3 \times 10^9
 M_\odot$ to $\rm \sim 3 \times 10^{11} M_\odot$).  Our results show that the
 dust-corrected red envelope of galaxy evolution is well modelled by a starburst
 at redshift $\rm z_f=5$ followed thereafter by passive evolution.  We explore
 the evolution of the ERG luminosity function (LF) from redshifts $\rm \langle
 z_{phot} \rangle =1.0$ to $\rm \langle z_{phot} \rangle =2.5$ and compare it
 with the global $\rm K_s$-band LF at redshifts $\rm 1<z_{phot}<2$. We find that
 the bright end of the ERG LF does not decrease from redshifts $\rm \langle
 z_{phot} \rangle =2.0$ to $\rm \langle z_{phot} \rangle =2.5$ and we connect
 this fact with the presence of progenitors of the local $\rm L>L^\ast$
 population at redshifts $\rm z_{phot} >2$. We determine lower limits of $\rm \rho_c=(6.1\pm 1.9) \times 10^{-5}\, Mpc^{-3}$ and $\rm \rho_c=(2.1\pm 1.1) \times 10^{-5}\, Mpc^{-3}$ on the comoving densities of progenitors of local massive  galaxies already assembled at redshifts $\rm
\langle z_{phot} \rangle=2.5$ and $\rm
\langle z_{phot} \rangle=3.5$, respectively. We have investigated the existence of high-redshift Lyman break galaxies massive enough to be included in this ERG sample. Out of an initial list of 
12 potential very high redshift candidates, we have identified 2 ERGs which 
have a high probability of lying at $\rm z_{phot}>4$. We discuss
 the advantages of multi-colour to single-colour selection techniques in
 obtaining reliable lists of very high-redshift candidate sources, and present revised lower redshift estimates for sources previously claimed 
as potential $\rm z > 5$ dropouts in recent studies.

\end{abstract}

\begin{keywords}
galaxies: evolution -- galaxies: formation -- galaxies: high-redshift
\end{keywords}

\setcounter{figure}{0}

\section{Introduction}
\label{sec-intro}

\parskip=0pt
 
     The study of `extremely red galaxies' (ERGs) is of importance for setting
constraints on the first epoch of massive galaxy formation, for exploring the
population of dust-reddened starbursts, and for testing models of galaxy
formation in general.  Usually defined as objects with $\rm R-K>5$ or $\rm
I-K>4$ (Vega), these galaxies have been widely studied since their discovery by
Elston, Rieke \& Rieke (1988). The traditional picture identifies the ERGs 
either with passively evolving  elliptical galaxies, 
or with young
starbursts strongly reddened by dust. This mixture has been confirmed by
spectroscopic (Dunlop et al. 1996; Soifer et al.1999; Cimatti et al. 2002;
Saracco et al. 2003) and sub-millimetre surveys (Cimatti et al. 1998; Dey et
al. 1999; Wehner, Barger \& Kneib 2002). However, the latest photometric and
morphological studies of these objects indicate that the ERG population is
perhaps more complex than previously considered (Smail et al. 2002; Miyazaki et
al. 2002; Yan \& Thompson 2003; Moustakas et al. 2004).

     Most previous samples of ERGs have been selected to a limiting magnitude of
$\rm K_s=20$ or brighter, and their inferred redshift distributions peak at $\rm
z \sim 1-2$ . New deeper observations reaching $\rm K_s=22$ and beyond (Maihara
et al. 2001, Saracco et al. 2001) are uncovering ERGs at higher redshifts. In
particular, a population of galaxies with very red near-infrared colours $\rm
J-K_s>3$ has been discovered to lie at redshifts $\rm z>2$ (Totani et al. 2001).
Saracco et al.(2004) claim that three massive galaxies of this kind in the
Hubble Deep Field South (HDFS) at $\rm \langle z_{phot} \rangle=2.7$ account for
about $\rm 40\%$ of the comoving density of the local early-type $\rm L>L^\ast$
galaxies.  However, these studies are based on small area surveys which cover a
few square arcmin of the sky.  Deep $\rm K_s$-band observations in wider areas
are required to obtain significant samples of high redshift ERGs and to
determine more accurate lower limits for the fraction of massive galaxies
already assembled at different redshifts.

\begin{figure}
\begin{center}
\includegraphics[width=1.0\hsize,height=0.95\hsize,angle=0] {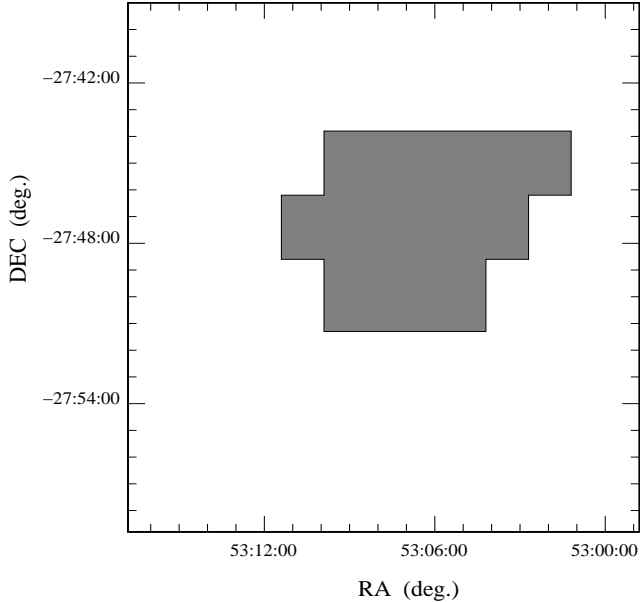}
\caption[]{\label{isaac} Schematic diagram of the sub-region covered by deep
VLT/ISAAC near-infrared imaging (shaded area) within the larger GOODS/CDFS field
with full HST-ACS coverage.}
\end{center}  
\end{figure}

The Great Observatories Origins Deep Survey (GOODS) (Dickinson et al. 2003) is
providing unprecedented multiwavelength data in $\sim$ 320 arcmin$^2$ centred on
the Chandra Deep Field South (CDFS) and Hubble Deep Field North (HDFN). Within
the GOODS/CDFS field, Roche, Dunlop \& Almaini (2003) selected 198 ERGs with $\rm K_s<22$ (Vega) and $\rm I_{775}-K_s>3.92$ from the 50.4 arcmin$^2$ for which deep near infra-red (near-IR) data have been obtained with the Infrared Spectrometer and Array Camera (ISAAC) on the `Antu' Very Large Telescope (Antu-VLT). This is the deepest significant sample of ERGs selected to date, and is the subject of the present study.  The region covered by the ISAAC observations within the
GOODS/CDFS field is shown in fig.\ref{isaac}. Other previous studies of ERGs
reaching $\rm K_s \sim 20$ exist in the CDFS. For example, the K20 survey has
yielded spectroscopic redshifts for ERGs in 32.2 arcmin$^2$ of this field
(Cimatti et al. 2002). More recently, the GOODS team (Moustakas et al. 2004)
selected a sample of ERGs from 163 arcmin$^2$ of the GOODS/CDFS field with data
from the Advanced Camera for Surveys (ACS) on board the Hubble Space Telescope
(HST). The area covered by this brighter ERG sample is three times greater than
that covered by the Roche et al. sample studied here. However, the Moustakas et
al. sample only reaches a depth of $\rm K_s \sim 20.2$ (equivalent to $\rm
K_s=22, AB$), and so the present study probes a different (and complementary)
region of parameter space to that already explored by the GOODS team.

Roche et al. (2003) studied the number density of ERGs within their $\rm K_s=22$
sample and compared the results with the predictions of different galaxy
formation models. They also demonstrated that the strong clustering displayed by
the ERG population extends to the faintest $\rm K_s$ magnitudes. In this work,
we present photometric redshift estimates for the ERGs in the Roche et
al. sample and explore the resulting implications for the nature and evolution
of ERGs.

The layout of this paper is as follows. First, in Section 2, we briefly review
the sample selection and explain the details of the multiwavelength
photometry. In Section 3 we discuss the photometric redshift techniques applied,
and present examples of individual redshift determinations. Then, in Section 4
we present our results and discuss raw and dust-corrected Hubble diagrams ($\rm
K_s-z_{phot}$), the red envelope of galaxy evolution, the evolution of the ERG
luminosity function (LF), and derived lower limits on the comoving densities of
progenitors of local $\rm L>L\ast$ galaxies. We  discuss the individual
properties of each of our high-redshift ($\rm z_{phot}>4$) candidate sources and revise the nature of those included in the Hubble Ultra Deep Field (UDF), in the light of the new publicly available ultra-deep images taken with the HST-ACS and the HST-Near Infrared Camera and Multi Object Spectrometer (NICMOS). We also
review the existence of an ERG cluster reported by Roche et al. (2003). Finally,
we present some concluding remarks in Section 5. We adopt throughout a cosmology
with $\rm H_o=70 \,{\rm km \, s^{-1} Mpc^{-1}}$, $\rm \Omega_M=0.3$ and $\rm
\Omega_\Lambda=0.7$.

\section{The sample-Multiwavelength photometry}
\label{sec-phot}

\begin{figure*}
\begin{center}
\includegraphics[width=0.8\hsize,angle=0] {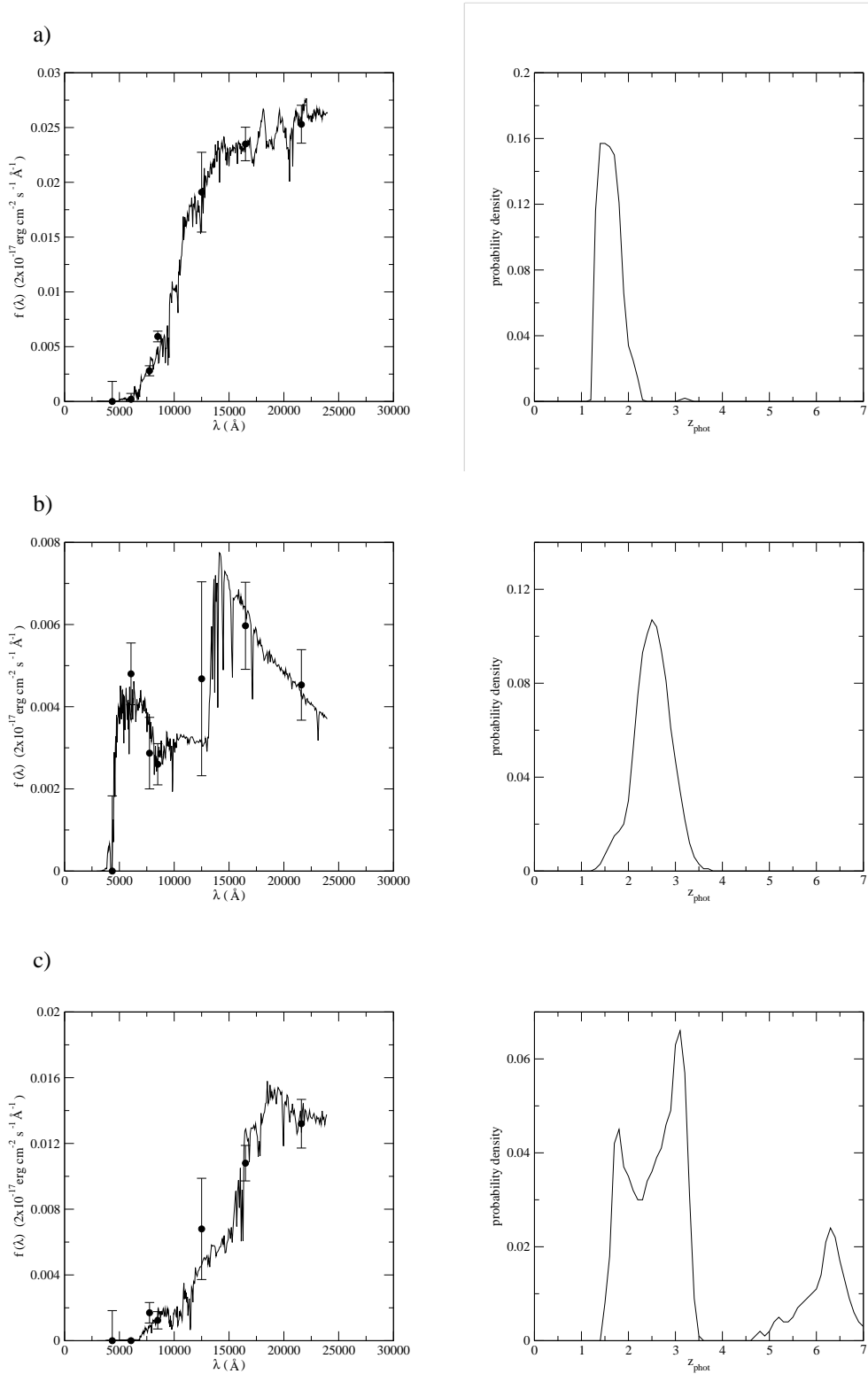}
\caption[]{ \label{fig-sedcomp} Example spectral energy distributions (left-hand
plots) and redshift probability density distributions (right-hand plots) for
three ERGs in the Roche et al. sample. a) an ERG with `hyperz' primary solution
$\rm z_{phot}=1.40$, b) an ERG with primary solution $\rm z_{phot}=2.52$, c) an
ERG with primary solution $\rm z_{phot}=3.10$.  The circles in the SED plots
correspond to the measured aperture magnitudes in each filter. The solid-line
curves indicate the best-fit template in each case.}
\end{center}  
\end{figure*}

\parskip=0pt

GOODS observations include optical and near-IR imaging in the B, V, I$\rm
_{775}$ and z bands with the ACS/HST and J, H and $\rm K_s$ bands with the
ISAAC-VLT. Roche et al. (2003) used ISAAC K$_s$-band data and ACS $\rm
I_{775}$-band data to select a sample of 198 ERGs with $\rm K_s<22$ (Vega) and
$\rm I_{775}-K_s>3.92$.  They performed the source extraction in the $\rm K_s$
band and photometry in the $\rm I_{775}$, J, H and K$_s$ bands using the public
code `SExtractor' (Bertin \& Arnouts, 1996). All the selected ERGs are at least
$\rm 3 \sigma$ detections on the $\rm K_s$-band images.  Further details of
their photometric measurements from the ground-based data and their preliminary
measurements on the $\rm I_{775}$-band ACS images are given in their paper. The
aperture magnitudes measured on ISAAC images by Roche et al. (2003) and used
here correspond to $2^{\prime \prime}$-diameter in all cases.

We performed photometric measurements for the four ACS bands on the stacked
GOODS 5-epoch images (V1.0 HST/ACS data release), using the corresponding
weighting maps. We looked for counterparts of the $\rm K_s$-selected ERGs on the
I$\rm_{775}$ and z-band images within an angular radius of $1^{\prime
\prime}$. For B and V bands we restricted the search of counterparts to
$0.5^{\prime \prime}$ to minimise the presence of interlopers. We also performed
the magnitude measurements using the public code `SExtractor'. Many ERGs
appeared as non-detected in the B and V bands, and some of them as non-detected
in the $\rm I_{775}$ or z bands. The algorithms for computing photometric
redshifts depend critically on the adopted value of limiting magnitudes in the
case of non detections, and a careful treatment of potential `dropouts' is
necessary to prevent the photometric redshift algorithms from finding fake high
redshift sources.  Therefore, in each case of a `SExtractor' non-detection, we
used the IRAF task `phot' to measure aperture magnitudes centred at the ERG $\rm
K_s$-band position.  We measured $2^{\prime \prime}$-diameter aperture
magnitudes on each ACS filter for those ERGs with $\rm z_{2^{\prime \prime}}< 26
\, (AB)$. Beyond $\rm z_{2^{\prime \prime}}= 26 \, (AB)$, we found that little
flux was missed using $1^{\prime \prime}$ instead of $2^{\prime
\prime}$-diameter aperture magnitudes. Thus, for objects with $\rm z_{2^{\prime
\prime}}>26 \,(AB)$, we used $1^{\prime \prime}$-diameter aperture magnitudes to
reduce the random errors, and corrected the values systematically using the
average offset between the $2^{\prime \prime}$ and $1^{\prime \prime}$-diameter
aperture magnitudes measured in each $\rm \Delta = 1$ magnitude bin.  After this
procedure, only a few objects remained as formally non-detected in the V or
redder ACS bands.

The photometry of three ERGs included in an initial list of potential very high redshift sources has been revised on the new available ACS and NICMOS ultra deep optical and near-IR images of the Hubble UDF (cf. Section 4.6). The procedure we followed to perform photometric measurements on these images is similar to the one used for the 5-epoch stacked GOODS HST/ACS images.

\section{Redshift estimations}
\label{sec-redestim}

\parskip=0pt

We computed photometric redshifts for the ERGs in the Roche et al. (2003) sample
with the public code `hyperz' (Bolzonella, Miralles \& Pell\'{o}, 2000), using
the seven passbands described in Section 2 (B, V, I$\rm_{775}$, z, J, H and
K$\rm_s$) and the GISSEL98 library of Bruzual \& Charlot (1993). The code
`hyperz' computes the probability of an object being at a given redshift $\rm
z_{phot}$ by searching for the best ($\chi^2$) fit to its photometric spectral
energy distribution (SED) provided by any of the template spectra available in
the library.  The redshift which yields the highest probability is known as the
`hyperz' primary solution and, in the following, we refer to the best-fit SED as
the one corresponding to this solution. Although this primary solution is the
one used in most cases, `hyperz' also allows one to construct a probability
density distribution in redshift space for each object.  This is very useful for
revealing possible degeneracies in the redshift determination, and for assessing
the significance of the primary solution. To account for dust obscuration, we
applied a Calzetti et al. (2000) reddening law, allowing the extinction in the
V-band (A$\rm_V$) to vary between 0 and 1. For the reddest objects, we ran
`hyperz' again allowing A$\rm_V$ to vary between 0 and 3. This did not produce
any substantial changes in the redshift estimations, but does have some impact
on the estimated absolute magnitudes and ages.

We also used the public `Bayesian photometric redshifts' (BPZ) code produced by
Ben\'{\i}tez (2000) to obtain a second, independent set of redshift estimates
for the ERGs in our sample. This code computes photometric redshifts using a
Bayesian approach (and also has the option of using a maximum likelihood
technique like the one used by `hyperz'). The agreement between the redshifts
estimated by `hyperz' and the BPZ code is very good in most cases. Accordingly,
with a few exceptions discussed individually in later sections, we have adopted
the `hyperz' results for the ERG sample simply because it provides not only the
redshift probability distribution, but also several other useful quantities in
the output (e.g.  k-corrected absolute magnitudes in a selected filter, V-band
dust extinction and age).

\begin{figure}
\begin{center}
\includegraphics[width=1.0\hsize,angle=0] {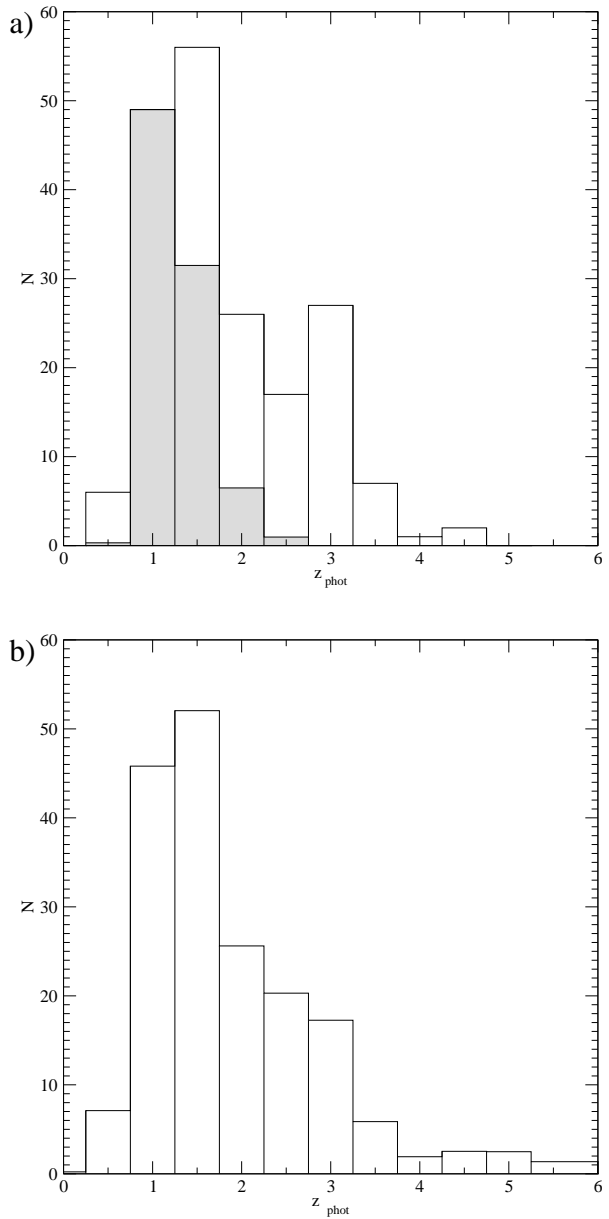}
\caption[]{\label{histo} Redshift distribution of the ERGs in Roche et al.
sample: a) taking into account only `hyperz' primary solutions, b) taking into
account the probability density distribution of each ERG in the redshift
space and small corrections due to the incompleteness of the sample. In the upper plot a), the redshift distribution of the shallower sample
of ERGs selected by Moustakas et al. (2004) in a wider area of the GOODS/CDFS
has been added for comparison (shaded histogram), after division by a factor of
3.08 in order to normalise their distribution to produce the same number of
objects in the redshift bin centred at $\rm z_{phot} =1$ as found in the present
study. }
\end{center}  
\end{figure}

Figure \ref{fig-sedcomp} shows examples of the best-fit SEDs (left-hand panels)
and redshift probability density distributions (right-hand panels) for three of
the ERGs in our sample. In the SED plots, the circles correspond to the measured
aperture magnitudes in each filter.  The solid-line curves indicate the best-fit
templates. The SED shown in fig. \ref{fig-sedcomp}-a) corresponds to an object with `hyperz' primary solution $\rm z_{phot}=1.40$. The accuracy in the fitting and the small error
bars in the photometry are reflected in a relatively small uncertainty in the
redshift estimation. The SED shown in fig. \ref{fig-sedcomp}-b) corresponds to an object with primary solution $\rm z_{phot}=2.52$. In this case, the larger error bars in the photometry produce a wider redshift probability density distribution.  Finally, the SED in fig. \ref{fig-sedcomp}-c) corresponds to an ERG with primary solution $\rm z_{phot}=3.10$. This is an example of an object with significant degeneracy in redshift space: the corresponding redshift distribution also shows non-negligible probabilities for this object to be either at $\rm z_{phot} \sim 1.8$ or at $\rm z_{phot}>5$.

\section{Results}

\subsection{The redshift distribution} 

Figure \ref{histo} shows our derived redshift distribution for Roche et
al. sample of ERGs. A few objects have been excluded either because their light
is contaminated by bright neighbours or because they are quite likely to be
stars given their very blue $\rm (J-K_s)$ colours. These objects are
characterised by very low probabilities of being at any redshift as deduced by
`hyperz'. We constructed the histogram in fig.  \ref{histo}-a) taking into
account `hyperz' primary solutions only, for ease of comparison with the Hubble
diagrams shown in Section 4.2. For the ERGs with identification numbers e1504
and e1605 in the Roche et al.  sample, we adopted the redshifts estimated by the
BPZ code by Ben\'{\i}tez (2000) instead of the `hyperz' primary solution for
reasons which are explained in Section 4.6. This histogram, as well as all the results presented hereafter,  include the revised redshifts of the potential high-z candidate ERGs e778, e1113 and e1272, as they were determined after measuring the photometry of these objects on the  NICMOS and ACS ultra deep images of the Hubble UDF.  We also show in fig.  \ref{histo}-a) the redshift
distribution for the shallower sample of ERGs in the HST-ACS field of the
GOODS/CDFS obtained by Moustakas et al. (2004) (shaded histogram),
 after division by a factor of 3.08 in order to normalise
their distribution to produce the same number of objects in the redshift bin
centred at $\rm z_{phot} =1$ as found in the present study.

The redshift distribution for the Roche et al. sample spans the range $\rm
z_{phot} \sim 0.5-4.75$. Moustakas et al. sample, which is approximately two
magnitudes shallower, only includes objects up to redshift $\rm
z_{phot}=2.5$. The maximum of our redshift distribution of $\rm K_s<22$ ERGs is
located at redshift $\rm z_{phot} \sim 1.5$.  We find a secondary maximum at
redshift $\rm z_{phot} \sim 3$. However, this secondary peak is not
statistically significant and disappears when probability densities are used to
construct the redshift distribution as shown in fig. \ref{histo}-b). This 
latter histogram also includes very small corrections for the incompleteness of the sample\footnote{Roche et al. ERG sample is considered to be $\sim 100\%$ and $\sim 80\%$ complete to $\rm K_s \approx 21.5$ and at $\rm 21.5 < K_s < 22.0$, respectively.} and, thus, should be taken as a more realistic representation of the redshift distribution of the ERGs in the GOODS/CDFS deep ISAAC field.

\begin{figure*} 
\begin{center}
\includegraphics[width=1.0\hsize,angle=0] {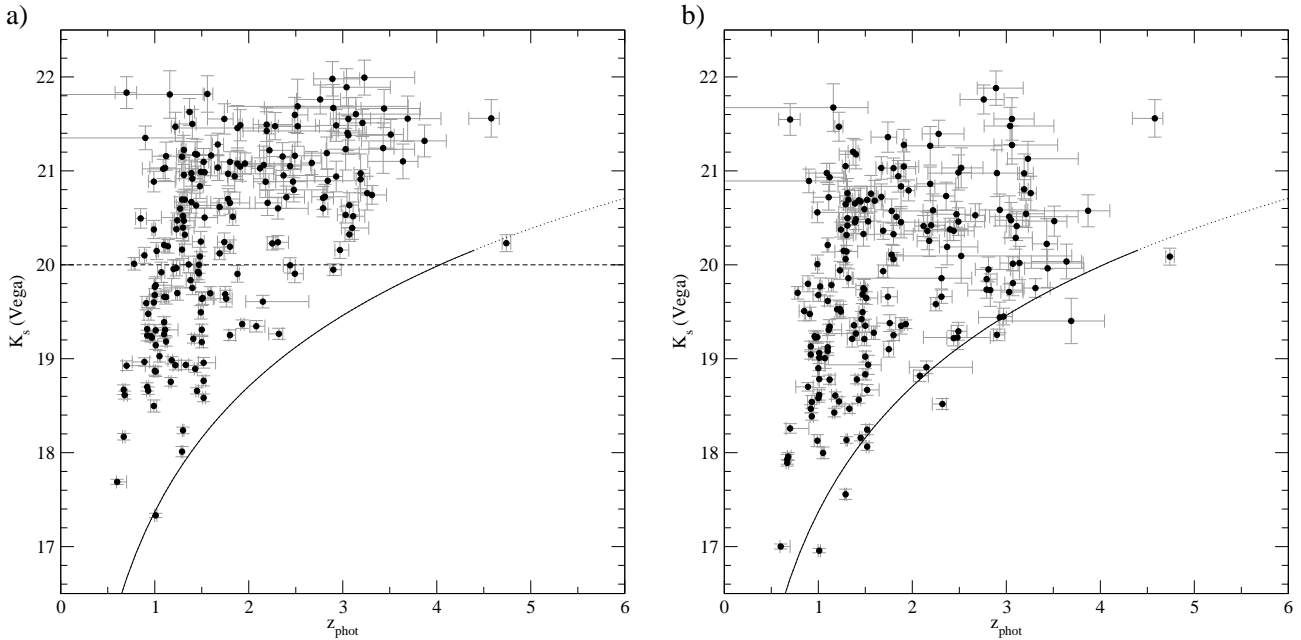}
\caption[]{\label{fig-Kvsz} Hubble diagram (total $\rm K_s$ magnitude versus
photometric redshifts) for the ERGs in the Roche et al.  sample. a) Original
$\rm K_s$ magnitudes; b) dust-corrected $\rm K_s$ magnitudes. The circles
correspond to `hyperz' primary solutions.  The error bars indicate 1$\sigma$
confidence levels. The dashed line on a) delimits the region of the Hubble
diagram to which $\rm K_s=20$ surveys have access.  The filled curve corresponds
to the empirical K-z relation for massive radio galaxies obtained by Willott et
al. (2003), which approximately corresponds to the passive evolution of a $\rm
3L^\ast$ starburst formed at redshift $\rm z_f=10$. The dotted line is a nominal
extrapolation of the same law.}
\end{center}  
\end{figure*}

\subsection{The Hubble diagram: $\rm K_s$ vs. $\rm z_{phot}$}

Figure \ref{fig-Kvsz} shows the Hubble diagram ($\rm K_s$ vs. $\rm z_{phot}$)
for the ERGs in the Roche et al. sample, corresponding to the redshift
distribution presented in fig. \ref{histo}.  The plots labelled as a) and b)
show the raw and dust-corrected $\rm K_s$ magnitudes as a function of
redshift, respectively. Total $\rm K_s$ magnitudes are considered in this case. We computed the dust-corrected $\rm K_s$ magnitudes
using the V-band extinction value $\rm A_V$ applied to the best-fit SED in
`hyperz'.  The rest-frame wavelength sampled by the $\rm K_s$-band is

\begin{equation}
\lambda_{\rm rf}=\frac{\lambda_{\rm K_s}}{1+\rm z_{phot}},
\end{equation}

\noindent where $\lambda_{\rm K_s}$ is the effective wavelength of the $\rm K_s$
filter, i.e. $\rm \lambda_{K_s}=2.16 \, \mu m$. In a dust-screen model, the extinction at a wavelength $\lambda_{\rm rf}$ is related to the extinction in the V-band ($\rm A_V$) by

\begin{equation}
\label{ext}
 \rm A_{\lambda_{\rm rf}}=\frac{k(\lambda_{\rm rf}) \, A_V}{R_V}.
\end{equation}

\noindent For the Calzetti et al. (2000) reddening law, $\rm R_V=4.05 \pm
0.80$ and $\rm k(\lambda_{\rm rf})$ is a power law in $\lambda_{\rm rf}$
(cf. `hyperz' user's manual). The dust-corrected $\rm K_s$ magnitude for each
source is given by the difference $\rm (K_s - \rm A_{\lambda_{\rm rf}})$. It is important to note that we assume the validity of the dust-screen model and the Calzetti et al. (2000) reddening law for modelling the extinction observed in ERGs.  We explored the use of other reddening laws in `hyperz' and found that  both  Milky Way and  Large Magellanic Cloud-type laws produce very similar raw and dust-corrected Hubble diagrams for the ERGs in Roche et al. sample. The study of other geometries for the distribution of dust is beyond the scope of this paper, but one should be  aware that they could change the relation given by eq. (\ref{ext}) between the extinction $\rm A_{\lambda_{\rm rf}}$ and the extinction in the V-band $\rm A_V$. This might have some impact on the derived properties of the most highly extincted ERGs. 

 In both plots \ref{fig-Kvsz}-a) and \ref{fig-Kvsz}-b), the circles indicate
`hyperz' primary solutions. The error bars correspond to 1$\sigma$ confidence
levels.  The overlaid solid curve shows the empirical K-z relation fitted for
massive radiogalaxies (Willott et al. 2003), which approximately corresponds to
the locus of passively evolving present-day $\rm 3L^\ast$ galaxies formed
instantaneously at redshift $\rm z_f=10$. The dotted line is a nominal
extrapolation of the same law. In fig. \ref{fig-Kvsz}-a), we have used a dashed
line to indicate the region of the Hubble diagram to which the surveys limited
at $\rm K_s=20$ have access.  Up to such a magnitude, only sources with $\rm
z_{phot} < 3$ are detected within our survey.  For a limiting magnitude $\rm
K_s=22$, the ERGs span the redshift interval $\rm z_{phot} \sim 0.5-4.75$ and
display a large dispersion in the $\rm K_s$ vs. $\rm z_{phot}$ relation.  In the
raw $\rm K_s$-magnitude Hubble diagram there is an obvious lack of objects near
the radio-galaxy locus beyond redshift $\rm z_{phot} \sim 3$. However,
comparison with fig. \ref{fig-Kvsz}-b) shows that this effect is a consequence
of the presence of dust, rather than indicating a real absence of intrinsically
bright, high-redshift ERGs.

The position of a galaxy in the dust-corrected $\rm K_s-z_{phot}$ diagram
depends, of course, on its age as well as its mass.  The radio-galaxy K-z
relation can be interpreted as indicating the behaviour of the highest mass
galaxies formed at very high redshift. However, galaxies may appear to be as
bright or brighter than high-redshift radio galaxies without being as massive
provided their stellar populations are sufficiently young and bright.  However,
as we show in Section 4.5, most of the galaxies around the radio-galaxy line do
in fact appear to at least be massive enough to be the already assembled
progenitors of the local $\rm L>L^\ast$ population.  The comparison of
fig. \ref{fig-Kvsz}-a) and \ref{fig-Kvsz}-b) also shows that the dispersion in
the $\rm K_s-z_{phot}$ relation for ERGs is not an effect produced by the
presence of different amounts of dust. We see an even larger dispersion for the
ERGs in the dust-corrected Hubble diagram.  This fact indicates that the ERG
population comprises objects spanning a wide range in mass.

To obtain an estimate of the minimum mass of each ERG, we used the $\rm
(k+e)$-corrected $\rm K_s$-band absolute magnitude of each source and computed a
lower limit for the luminosity the galaxy would have at redshift $\rm z=0$,
assuming passive evolution. We explain the details of the calculation of the
$\rm (k+e)$-corrected $\rm K_s$-band absolute magnitudes in Section 4.5. The
luminosity L and the absolute magnitude M in a given band are related by the
relation $\rm L/L^\ast=10^{-0.4(M-M^\ast)}$, where we considered $\rm
M_{K_s}^\ast=-24.2$ for h=0.7, from the 2dF local luminosity function (Cole et
al. 2001).  Although the exact mass-to-light ratio depends on the assumed
initial mass function (IMF) and the age of each galaxy, we estimate the mass of
each ERG as $\rm (L/L^\ast)\times 10^{11}M_\odot$, where L is the luminosity the
galaxy would have today at redshift $\rm z=0$.  Estimating the masses in this
way, we find that the ERGs in the Roche et al. sample span two decades in mass,
from $\rm \sim 3 \times 10^9 M_\odot$ to $\rm \sim 3 \times 10^{11} M_\odot$.

\subsection{The red envelope of galaxy evolution}

\parskip=0pt
 
\begin{figure*}
\begin{center}
\includegraphics[width=1.0\hsize,angle=0] {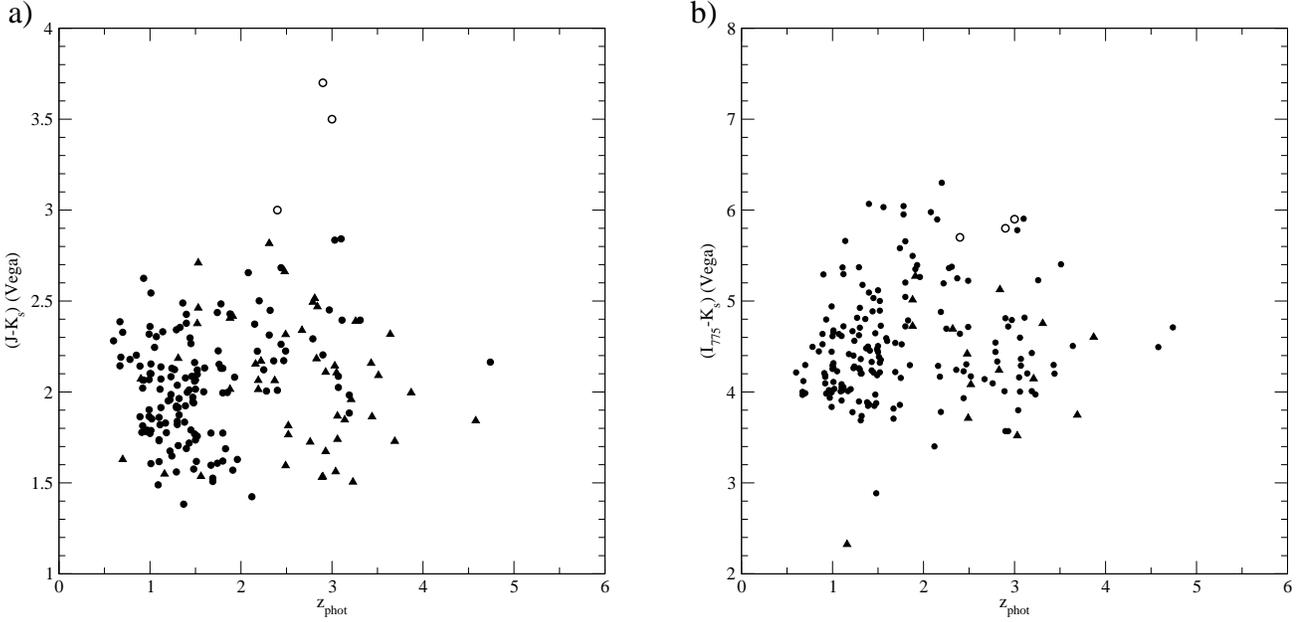}
\caption[]{\label{colz} $\rm (J-K_s)$ (a) and $\rm (I_{775}-K_s)$ colours (b)
vs. photometric redshifts.  The filled circles correspond to exact values and
the filled up-triangles, to lower limits for the sources with error $\rm
\varepsilon >0.5$ in the J or $\rm I_{775}$ aperture magnitudes. The open
circles indicate the colours of the three high redshift massive galaxies in the
HDFS reported by Saracco et al. (2004).}
\end{center}  
\end{figure*}

Figures \ref{colz}-a) and \ref{colz}-b) show the $\rm (J-K_s)$ and $\rm
(I_{775}-K_s)$ colours, respectively, as a function of photometric redshift for
the ERGs in the Roche et al.  sample. Only `hyperz' primary solutions are
considered in this case.  In fig. \ref{colz}-a), the filled circles correspond
to those objects with J-band $2^{\prime \prime}$-diameter aperture magnitude
above the $\rm 2\sigma$-confidence limit, in this case $\rm J\leq J_{lim.}=23.5$
(Vega). The up-triangles indicate lower limits to the colours of those objects
with $\rm J>23.5$, computed as $\rm (23.5 - \rm K_s)$. In fig. \ref{colz}-b),
the mixture of techniques we applied to measure aperture magnitudes on ACS
images, i.e. `SExtractor' plus IRAF `phot', does not allow us to set the
2$\sigma$ confidence limit at a given $\rm I_{775}$-band magnitude. Thus, we
considered as precise colours (filled circles) those with $\rm I_{775}$-band
magnitude error $\rm \varepsilon <0.5$. For objects with $\rm I_{775}$-band
magnitude error $\rm \varepsilon >0.5$ we computed lower limits for the colours
as $\rm (I_{775}-\varepsilon-K_s)$ (up-triangles).  For comparison, we have
added in both plots, \ref{colz}-a) and \ref{colz}-b), the colours of the three
massive galaxies in the HDFS at redshift $\rm z_{phot} \geq 2.4$, reported by
Saracco et al. (2004) (open circles).
                
\begin{figure*}
\begin{center}
\includegraphics[width=1.0\hsize,angle=0] {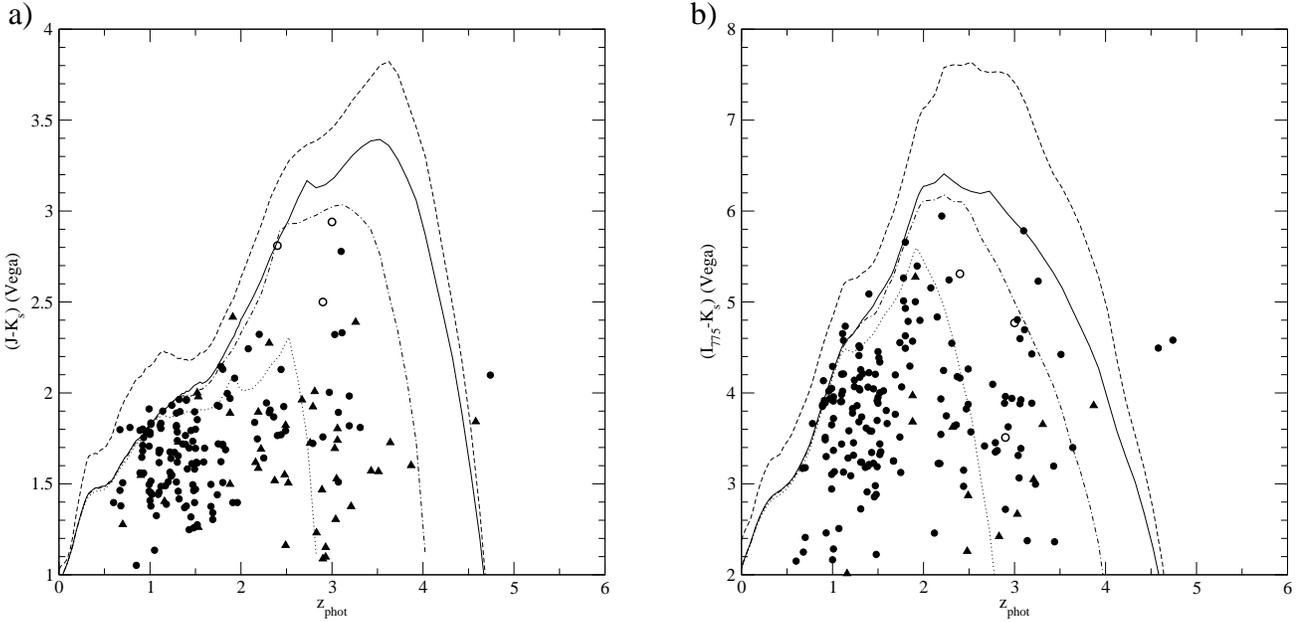}
\caption[]{\label{coldustz} Dust-corrected $\rm (J-K_s)$ (a) and $\rm
(I_{775}-K_s)$ colours (b) vs. photometric redshifts. The labels for the filled
circles, filled up-triangles and open circles are the same as in
fig. \ref{colz}. The different lines show the expected colours for  starbursts at different redshifts with passive evolution thereafter and metallicity $\rm
Z=Z_\odot$: $\rm z_f=5$ (solid), $\rm z_f=4$ (dashed-dotted) and $\rm z_f=3$ (dotted)  . The dashed line corresponds to a similar starburst at redshift $\rm z_f=5$, but with a higher
metallicity $\rm Z=2.5 \, Z_\odot$.}
\end{center}  
\end{figure*}
   
The extremely red colours observed in ERGs are due to two coupled factors: age
and dust. As our aim is to study which evolutionary model produces the reddest
colours observed in the ERGs, we need to correct for the dust effects in order
to separate the contribution to the red colours of the evolved stellar
populations.  Figures \ref{coldustz}-a) and \ref{coldustz}-b) show the
dust-corrected $\rm (J-K_s)$ and $\rm (I_{775}-K_s)$ colours, respectively, as a
function of redshift. We computed the dust-corrections for the colours
subtracting from each magnitude the corresponding extinction value given by
eq. (\ref{ext}). The symbol labels in both plots of fig. \ref{coldustz} are
equivalent to those in fig. \ref{colz}. We used the public code `GALAXEV'
(Bruzual and Charlot 2003) to compute the expected $\rm (J-K_s)$ and $\rm
(I_{775}-K_s)$ colours of different kinds of dust-free passively evolving
galaxies, in order to obtain a suitable model for the red envelope delimited by
the ERG largest colours. We found that an instantaneous starburst at redshift
$\rm z_f=5$ with passive evolution thereafter and solar metallicity (solid lines
in fig.\ref{coldustz}) models quite well the dust-corrected red envelope. A
similar starburst formed at redshift $\rm z_f=5$, but with a higher metallicity
$\rm Z=2.5 \, Z_\odot$ (dashed line), has been added for comparison. The plots
in fig. \ref{coldustz} show that solar metallicity $\rm Z=Z_\odot$ is sufficient
to explain the reddest colours observed. Even the few apparent outliers at
redshift $\rm z_{phot} \sim 0.5-2.0$ can be explained by such a model once the
error bars are taken into account, which are $\rm \sim 0.2 \, mag$ for these
objects.  On the other hand, the modelling of the expected colours for starbursts at different redshifts allows us to investigate the epoch since when the ERGs lying at redshifts $\rm z_{phot}\gsim 2$  have been following  passive evolution. The dotted and dashed-dotted lines in fig. \ref{coldustz} correspond to the modelled colours for starbursts at redshifts $\rm z_f=3$ and  $\rm z_f=4$, respectively (both  with passive evolution thereafter and solar metallicity $\rm Z=Z_\odot$). We see that the reddest  ERGs  lying at redshifts $\rm z_{phot}\sim 2 \, (3)$  have been passively evolving since redshift $\rm z_{phot} \sim 3 \, (4)$. This does not necessarily mean that these objects have been formed in these epochs, but corresponds to the formation redshifts of the youngest stellar population present in them.

\begin{figure}
\begin{center}
\includegraphics[width=1.0\hsize,angle=0] {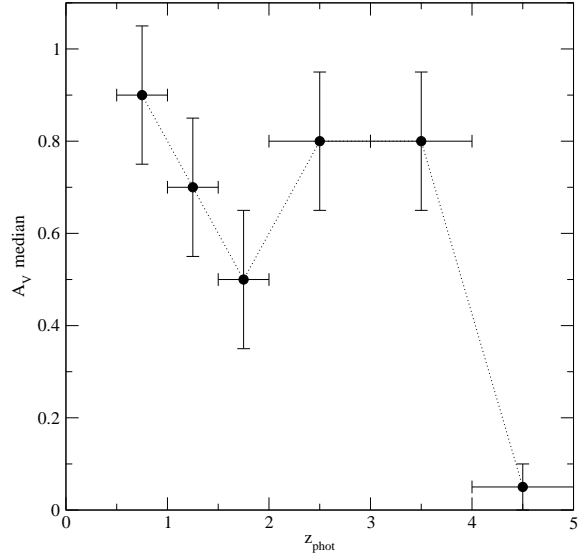}
\caption[]{\label{Avz} Median of the V-band extinction values applied to the best-fit SEDs vs. photometric redshifts. Only 'hyperz' primary solutions are taken into account. The horizontal  and vertical error bars indicate the binning in  redshift space and the largest binning used in `hyperz' for the iteration through  the $\rm A_V$ values, respectively. }
\end{center}  
\end{figure}

The comparison of fig. \ref{colz} and fig. \ref{coldustz} provides a
simple and useful way to understand the extent to which the presence of dust is
responsible for the original extremely red colours of ERGs. This fact is summarised in fig. \ref{Avz}, where we show the median of the V-band  extinction values $\rm A_V$ necessary to deredden the ERG best-fit SEDs at different redshifts. We see that considerable median extinction values ($\rm A_V \geq 0.5$) are necessary  to explain the red colours observed in ERGs up to redshift $\rm z_{phot} \sim 4$.  For objects below redshift  $\rm z_{phot} = 1$, the modelled $\rm A_V$ values are very large, illustrating that only extremely dusty starbursts can classify as ERGs at relatively low redshifts. The median of the extinction $\rm A_V$ is lower  at redshifts $\rm z_{phot} \sim 1 - 2$, where perhaps the essentially non-dusty evolved stellar populations are expected to make the most of their contribution. At redshifts $\rm z_{phot} \sim 2 - 4$, where according to fig. \ref{coldustz} most of the objects seem to have experienced recent starbursts, the median  of the extinction $\rm A_V$ increases again. For objects beyond redshift $\rm z_{phot}=4$, nearly no extinction is required indicating that, at this stage, the extremely red colours are mainly due to the shift of the Lyman break to optical wavelengths rather than dust.

\subsection{The evolving luminosity function of ERGs}

\begin{figure}
\begin{center}
\includegraphics[width=1.0\hsize,angle=0] {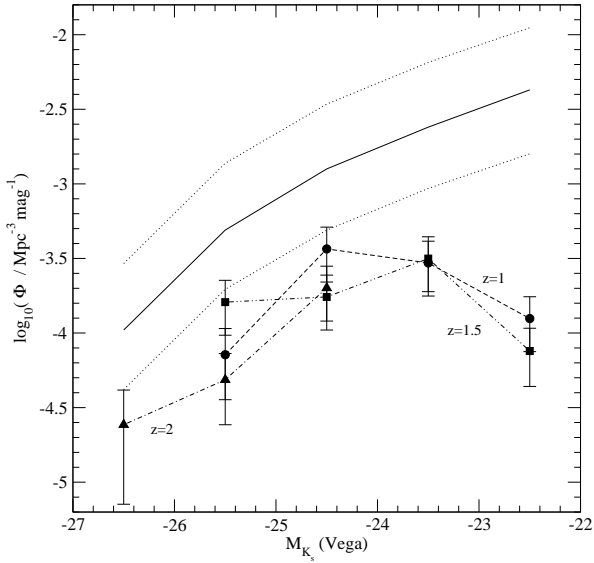}
\caption[]{\label{fig-LF12} Luminosity function of ERGs at different redshifts:
$\rm \langle z_{phot}\rangle=1.0$ (circles), $\rm \langle z_{phot}\rangle=1.5$
(squares) and $\rm \langle z_{phot}\rangle=2.0$ (up-triangles). The average
global $\rm K_s$-band LF in the HDFN and HDFS has been overlaid for comparison
(solid line).  The latter has been computed integrating in each absolute
magnitude bin the Schechter functions fitted by Bolzonella et al. (2002) in
these fields. The error bars have been estimated using the maximum uncertainty
in the normalisation parameter $\rm \phi^\ast$, obtained for the HDFS (dotted
lines).  }
\end{center}  
\end{figure}

\begin{figure}
\begin{center}
\includegraphics[width=1.0\hsize,angle=0] {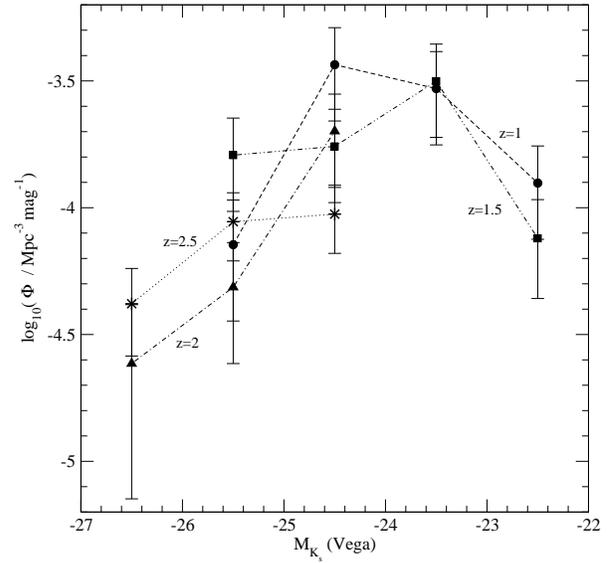}
\caption[]{\label{fig-LF23} Luminosity function of ERGs at different
redshifts. The references are the same as in fig. \ref{fig-LF12}. The LF at
redshifts $\rm 2<z_{phot}<3$ has been added (star-like symbols). This plot shows
that the bright end of the ERG LF does not obviously decrease from redshifts
$\rm \langle z_{phot} \rangle = 2.0$ to $\rm \langle z_{phot} \rangle = 2.5$.}
\end{center}  
\end{figure}

 Figures \ref{fig-LF12} and \ref{fig-LF23} show the luminosity function (LF) of
the ERGs in Roche et al. sample in different redshift intervals.  We assigned
to each object the $\rm K_s$-band k-corrected absolute magnitude ($\rm M_{K_s}$)
obtained as part of the `hyperz' output. We considered `hyperz' primary solutions
for binning the data in redshift space.  We computed the comoving densities per
absolute magnitude bin by dividing the number of sources in each bin by the
corresponding comoving volume. To take into account the limits of the survey
($\rm K_s=22$), we corrected the contribution of each source by a weighting
factor $\rm V_{maxbin}/V_{maxobs}$, where $\rm V_{maxbin}$ is the volume
determined by the maximum redshift of the bin and $\rm V_{maxobs}$ is the volume
corresponding to the maximum redshift at which the source would still be
included in the survey (provided it is lower than the maximum redshift of the
bin). We also applied a correction factor to account for a slight incompleteness of the sample at $\rm 21.5 < K_s < 22.0$.

    Fig. \ref{fig-LF12} shows the ERG LF at redshifts $\rm \langle z_{phot}
\rangle =1.0$, $\rm \langle z_{phot} \rangle =1.5$ and $\rm \langle z_{phot}
\rangle =2.0$, indicated by circles, squares and up-triangles, respectively. The
error bars correspond to the maximum of the Poissonian errors and the errors due
to cosmic variance, which we take on average as 40\% in the number counts at
these redshifts (cf. Somerville et al. 2003). The inclusion of the cosmic
variance is fundamental for highly-clustered populations, as ERGs at redshifts
$\rm z \sim 1-2$ are known to be.  The global $\rm K_s$-band LF at redshifts
$\rm 1<z_{phot}<2$ has been added for comparison. The solid line represents the
average $\rm K_s$-band LF of the HDFS and HDFN, obtained by integrating in each
absolute magnitude bin the best Schechter function fitted by Bolzonella, Pell\'o \& Maccagni (2002) for datasets in these fields up to a limiting magnitude of $\rm K_s \sim 23$. The dotted lines indicate the largest error bars in the normalisation parameter $\rm \phi^\ast$, corresponding to the fitting of the HDFS data. As expected, our ERG LF reproduces the shape of the bright end of the global $\rm K_s$-band LF. The differences at fainter magnitudes are at least in part due to the nature of the ERG colour selection. We find only 3 very luminous galaxies with $\rm -27<M_{K_s}<-26$ with estimated redshifts $\rm z_{phot} \, \epsilon$
[0.75;2.25], 2 of which have redshifts in the interval [1.75;2.25]. These 2
sources have $\rm (I_{775}-K_s)\geq 5.9$ and $\rm (J-K_s)> 2.3$.  For the
absolute magnitude range $\rm -26<M_{K_s}<-24$, we do not observe any evolution
in the ERG LF from redshifts $\rm \langle z_{phot} \rangle=1.0$ to $\rm \langle
z_{phot} \rangle=2.0$. Any fluctuation in the mean values of the LF in this
absolute magnitude bin can be accounted for within the cosmic variance error
bars.  For absolute magnitudes $\rm -24<M_{K_s}$ we do not observe any evolution
in the LF from redshifts $\rm \langle z_{phot} \rangle=1.0$ to $\rm \langle
z_{phot} \rangle=1.5$. The limits of the survey  do not allow us to explore the LF at redshift $\rm \langle z_{phot} \rangle=2.0$ for such faint objects.

In figure \ref{fig-LF23}, we show again the ERG LF at redshifts $\rm \langle
z_{phot}\rangle=1.0,1.5,2.0$ and we add the ERG LF at $\rm \langle
z_{phot}\rangle=2.5$, for comparison.  We computed the latter taking into
account all the objects in the redshift interval [2.0,3.0]. At these redshifts,
we have adopted a slightly lower value for the typical cosmic variance, only
30\% in the number of counts on average (Somerville et al. 2003). This plot
shows that the bright end of the ERG LF does not decrease significantly from
redshifts $\rm \langle z_{phot}\rangle=2.0$ to $\rm \langle
z_{phot}\rangle=2.5$.  This result confirms the existence of a population of
extremely red bright galaxies at high redshifts.  We find 7 objects with
estimated redshifts in the interval [2.0,3.0] and absolute magnitudes $\rm
-27<M_{K_s}<-26$, 6 of which have $\rm (I_{775}-K_s)\geq 4.8$, placing them
among the very reddest objects at these redshifts. Moreover, all of these
objects have $\rm (J-K_s) \geq 2.2$. In the next section we investigate the
change of the absolute magnitudes of ERGs after evolution to redshift $\rm z=0$
and explore the comoving densities of potential progenitors of the local $\rm
L>L^\ast$ population.

\subsection{Comoving densities of massive galaxies}

In this section we study the comoving densities of ERGs that will still
contribute to the bright end of the $\rm K_s$-band LF once they evolve down to
redshift $\rm z=0$. If we assume passive evolution for ERGs from the epoch of
observation, the maximum dimming of the absolute magnitude $\rm M_{K_s}$ can be
estimated and, thus, a lower limit to the luminosity each object would have at
redshift z=0 obtained. To do this, we need to estimate an evolutionary
correction (e-correction) factor at redshift $\rm z=0$ which, when added to the
`hyperz' k-corrected absolute magnitude, yields the absolute magnitude the galaxy would have today.  The e-correction for each galaxy depends on its spectral type and age.  We used the public code `GALAXEV' (Bruzual and Charlot 2003) to
compute the absolute magnitudes at redshift $\rm z=0$ of the spectral types
corresponding to our ERGs at different ages. The correction applied to each
object is the difference in the absolute magnitude the object would have today
and at the age it had when its light was emitted. The objects that are
considered to be progenitors of the local $\rm L>L^\ast$ population are those
with $\rm (k+e)$-corrected absolute magnitude $\rm M_{K_s}<M_{K_s}^\ast$, given
the relationship $\rm L/L^\ast=10^{-0.4(M-M^\ast)}$. In a passive evolution
scenario, these ERGs must have contained a minimum stellar mass of $\rm \sim
10^{11} \, M_{\odot}$ when their light was emitted.

\begin{figure}
\begin{center}
\includegraphics[width=1.0\hsize,angle=0] {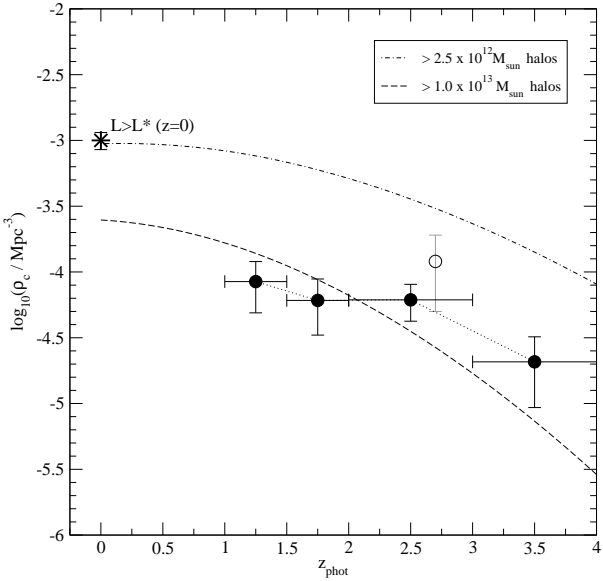}
\caption[]{\label{fig-Lcd} Lower limits on the comoving densities of ERGs progenitors
of local $\rm L>L^\ast$ galaxies  as a function of redshift (filled circles). The
horizontal error bars indicate the binning in the redshift space. The vertical
error bars correspond to Poissonian errors in the number of objects. The lower
limit on the comoving density of progenitors of $\rm L>L^\ast$ galaxies at
redshift $\rm \langle z_{phot} \rangle=2.7$ obtained by Saracco et al. (2004) (open circle) and the comoving density of local $\rm L>L^\ast$ galaxies (star-like symbol) have been added for comparison. The dashed-dotted  and dashed lines show the comoving densities of dark matter halos with masses $\rm M> 2.5 \times 10^{12}\, M_\odot$ and $\rm M> 1.0 \times 10^{13}\, M_\odot$, respectively, as obtained from $\rm \Lambda CDM$ models of structure formation.}
\end{center}  
\end{figure}

     Figure \ref{fig-Lcd} shows the comoving densities at different redshifts of
     the 27 ERGs in Roche et al. sample which are expected to have $\rm
     L>L^\ast$ at redshift $\rm z=0$ under passive evolution (filled circles).       We used $\rm
     M_{K_s}^\ast=-24.2$ for h=0.7 (Cole et al. 2001).   Once more, we computed
     the comoving densities dividing the number of sources in each redshift bin
     by the corresponding comoving volume. We divided the sample in bins of
     width $\rm \Delta z_{phot}=0.5$ up to redshift $\rm z_{phot}=2$ and $\rm
     \Delta z_{phot}=1$ for redshifts $\rm 2<z_{phot}<4$. To compute these
     comoving densities, we did not use `hyperz' primary solutions but a
     redshift probability density distribution that we constructed normalising
     for each source the percentage probabilities given in the `hyperz' output
     file \_log.phot. In this way, each source may contribute in a fraction to
     different redshift bins. Only sources with `hyperz' primary solutions $\rm
     z_{phot}<4$ have been taken into account. We applied a weighting factor of
     the kind $\rm V_{maxbin}/V_{maxobs}$, as it was explained in Section 4.4,
     to the normalised probability density distribution of each source. However,
     in each case, the weighting factor has been estimated using only the
     k-corrected absolute magnitude $\rm M_{K_s}$ given by `hyperz' in the
     output.  A more rigurous procedure should take into account a probability
     density distribution for the absolute magnitudes in correspondence with the
     probability density distribution for redshifts. We also corrected for the incompleteness of the sample. The error bars for the
     comoving densities correspond to Poissonian errors in the number of
     sources, which are dominant in this case. We added for comparison  the comoving density of the local $\rm L>L^\ast$ population (star-like symbol), which we computed as the average of the values obtained integrating the local $\rm K_s$-band LFs fitted by Cole et al. (2001) and Kochanek et al. (2001).

Our results show that there is no significant change in the comoving densities
of ERG progenitors of $\rm L>L^\ast$ galaxies between redshifts $\rm
1.5<z_{phot}<2.0$ and $\rm 2<z_{phot}<3$, which is consistent with the results
presented in Section 4.3 for the bright end of the LF. Actually, all the
galaxies with $\rm -27<M_{K_s}<-26$ in Section 4.3 are included in the
progenitors of $\rm L>L^\ast (z=0)$ subsample.  The values we obtain for the comoving densities are: $\rm \rho_c=(8.5\pm
3.5) \times 10^{-5}\, Mpc^{-3}$ for $\rm \langle z_{phot} \rangle=1.25$, $\rm
\rho_c=(6.1\pm 2.8) \times 10^{-5}\, Mpc^{-3}$ for $\rm \langle z_{phot}
\rangle=1.75$, $\rm \rho_c=(6.1\pm 1.9) \times 10^{-5}\, Mpc^{-3}$ for $\rm
\langle z_{phot} \rangle=2.5$ and $\rm
\rho_c=(2.1\pm 1.1) \times 10^{-5}\, Mpc^{-3}$ for $\rm \langle z_{phot}
\rangle=3.5$.  It is important to note that the relatively
small comoving densities at $\rm \langle z_{phot} \rangle=1.25$ and $\rm \langle
z_{phot} \rangle=1.75$ (more than 10 times smaller than the corresponding value we adopted for local $\rm L>L^\ast$ galaxies) could in part be  due to the ERG colour cutoff. If this is the case, we should conclude that the ERGs at redshifts $\rm z \sim 1-2$ cannot account for the whole population of progenitors of  the local massive galaxies.  Although
the ERGs reproduce quite well the shape of the bright end of the global $\rm K_s$-band LF, the values are lower, indicating that larger comoving densities of massive objects are expected to be obtained in a $\rm K_s$-selected sample without a colour selection.  Thus, the conclusion of a constant comoving density of massive objects from redshifts $\rm 1.5<z_{phot}<2.0$ and $\rm 2<z_{phot}<3$ is in
principle only applicable to ERGs, and may of course not hold for a general $\rm
K_s$-selected sample.  The mean value we obtain for the lower limit to the
comoving density of $\rm L>L^\ast$ progenitors at $\rm \langle z_{phot}
\rangle=2.5$ is only one half of the mean value obtained by Saracco et
al. (2004) at a similar redshift (open circle in fig. \ref{fig-Lcd}), but there
is no significant discrepancy between the results given the large error bars. We
conclude that Saracco et al. large mean comoving density might be due to a
simple cosmic variance effect and much larger samples of $\rm K_s$-selected
galaxies are necessary to establish a robust estimate of the fraction of the
stellar mass already assembled in ERGs at these redshifts. \footnote {Saracco et al. claim to reproduce about $40\%$ of the comoving density of the local  massive galaxies, after comparison with the local comoving density of $\rm L>L^\ast$ early-type galaxies obtained from the LF fitted by Marzke et al. (1998) using B-band data. Instead, we prefer to compare the comoving densities with the local values obtained integrating more recent  LFs directly  fitted on K-band data (Cole et al. 2001, Kochanek et al. 2001).}

In fig.  \ref{fig-Lcd} we show the  comoving densities of dark matter halos with masses  $\rm M> 1.0 \times 10^{13}\, M_\odot$ (dashed lines) at different redshifts, as they are predicted  by $\rm \Lambda CDM$ models of structure formation (Kauffmann \& Charlot 1998, Somerville \& Primack 1999). We also show  the  comoving densities of dark matter halos with masses  $\rm M> 2.5 \times 10^{12}\, M_\odot$ (dashed-dotted lines), a mass threshold deliberately selected to coincide with the comoving densities of $\rm L>L^\ast$ galaxies at $\rm z=0$. At any redshift, our comoving densities of progenitors of local massive galaxies are lower than the corresponding comoving densities of dark matter halos massive enough to host these galaxies. For instance, the comoving densities of $\rm M> 2.5 \times 10^{12}\, M_\odot$ halos decreases less than a factor $\sim 2$ from redshifts $\rm z=0$ to $\rm z=1.5$, while the comoving densities of ERGs progenitors of local massive galaxies are more than 5 times below this value. This leads to two extreme interpretations: i)  even taking into account the possible production of mergers (i.e. the increasing comoving densities of dark matter halos more massive than a given mass threshold with time), ERGs do not seem to be sufficient to account for all the progenitors of local $\rm L>L^\ast$  galaxies.  This would reinforce the idea that the ERG colour cutoff must be at least partially responsible for this deficiency, which appears to be more important at lower ($\rm z_{phot} \sim 1-2$) than higher ($\rm z_{phot} \sim 3-4$) redshifts. ii) ERGs do account for the progenitors of local massive galaxies but the passive evolution scenario is not completely  valid. Galaxies which are not sufficiently massive at a given redshift might build up more stellar mass at later epochs and, thus, should also be considered as progenitors of the local $\rm L>L^\ast$ galaxy population. The massive ERGs at $\rm z>2$ could instead be progenitors of much more massive ($\rm \sim 4 L^\ast$) local galaxies, in which case our results show that all of them are already in place at redshift $\rm z \sim 3.5$.

Interestingly, the progenitors of $\rm L>L^\ast$ galaxies are among the reddest
ERGs. At different redshifts, these massive galaxies in the Roche et al. sample
have $\rm \langle (I_{775}-K_s) \rangle \approx 4.8-5.1$.  The mean $\rm
(J-K_s)$ colours increase with redshift, as it is expected when the 4000$\rm
\AA$ break enters the region between the J and $\rm K_s$ filters.  In all cases,
$\rm (J-K_s)>2$.  Some of these massive galaxies are among the reddest objects
in $\rm (J-K_s)$ at redshifts $\rm z_{phot}>2$, due to the combined effects of
the break and considerable amounts of dust (we find V-band extinction values of
up to $\rm A_V=1.8$ for these massive objects). The characteristics of these
sources are consistent with the `hyper extremely red object' population
discovered by Totani et al.  (2001).  However, not all the $\rm
(J-K_s)$ reddest objects evolve to $\rm L>L^\ast$ sources in a passive evolution
scenario. In some cases, the extremely red $\rm (J-K_s)$ colours are almost
exclusively due to very large dust extinctions. Recently, some extremely red $\rm (J-K_s)$ sources have been found to be counterparts of sub-millimetre sources, which are known to be mainly located at redshifts $\rm
2<z<4$ (Dunlop 2001, Aretxaga et al. 2003, Chapman et al. 2003). However it is not yet clear whether the ERGs associated with sub-millimetre sources are primarily high or low mass objects (Frayer et al. 2004).

\subsection{Sources beyond redshift $\rm z=4$ - 
the ERG - Lyman-break connection}

In this section we focus on the properties of the Lyman break galaxies in the
GOODS/CDFS included in the Roche et al. ERG sample. 
These objects could be candidates
for massive galaxies at very high redshift.  Here we summarise the 
detailed properties of the individual objects and discuss the probability that they are genuinely
located at such high redshifts.
       
       Several studies have already been published on the selection of very high redshift candidate sources in the GOODS fields (Stanway, Bunker \& McMahon 2003; Bouwens  et al. 2003; Bremer et al. 2004; Dickinson et al. 2004). In most of these studies a colour cutoff selection technique has been applied
(usually $\rm (I_{775}-z)>1.3-1.5, AB$) to select candidates at redshifts $\rm z>5$. We find 9 of these candidate sources in the Roche et al. 
       ERG sample. However, we cannot confirm any of these sources as being at very high redshift. We argue that the fraction of low-mass star and lower-redshift ERG
       contaminants is usually under-estimated when a single-colour selection
       criterion is applied.  Multicolour photometry appears to be a much more
       powerful way to obtain reliable candidate lists of very high redshift
       objects, and allows the investigation of degerate solutions in 
       redshift space.

       Using 5-epoch GOODS HST-ACS (V1.0) and VLT-ISAAC data, we obtained an initial `hyperz' output list containing 12 sources with
       a primary solution at $\rm z_{phot}>4$, which adopted as candidates for
       detailed inspection. We present the properties of the complete list 
       of 12 candidate sources in Table 1. We used the public code BPZ  (Ben\'{\i}tez 2000) to obtain estimated redshifts for these objects in an independent
       way.  We also made an individual study of each of our 12 candidate
       sources, inspecting their ACS images and analysing their magnitudes and
       colours in different bands.  After a first analysis, we rejected the following 7/12 $\rm z>4$
       candidates for the reasons explained below:


\begin{itemize}
\item {\bf e114-e566-e967-e2386} : these sources appear to be at redshifts
$\rm z_{phot}>5.3$ from `hyperz' and all of them have $\rm (I_{775}-z)\sim 1.3-1.5, AB$. 
Only e566 is not confirmed by the BPZ code as a
very high redshift source (it is estimated to be at redshift $\rm
z_{phot}=1.28$). However, the probabilities associated with the `hyperz' primary
solutions, obtained from the $\rm \chi^2$ minimisation, are very low ($\rm
P<19\%$) in all cases.  e967 corresponds to source number 4 in Stanway et al. list 
and source number 8 in the list of Bremer et al. e114 and e2386 are sources number 13 and 9 in the 
Bremer et al. list, respectively. These sources are unresolved on ACS images and have
quite blue near-IR colours, $\rm (J-K_s)<1.7$ (Vega) in all cases. e967 and
e2386 are significant detections in the V-band ($\rm V=27.9 \pm 0.4$ and $\rm
V=27.4 \pm 0.3$, AB, respectively) and e566, in the B-band ($\rm B=23.7 \pm
0.1$, AB). Thus, it is more likely that they are cool stars than high redshift
galaxies. Indeed, e967 has been spectroscopically confirmed as such
(cf. Dickinson et al. 2004). These cool stars are known to be among the main
contaminants of very high redshift candidates.

\item {\bf e1504}: this source has `hyperz' primary solution $\rm
z_{phot}=4.63$, but actually has a probability $\rm P \equiv 0$ at any
redshift between 0 and 10. The BPZ code gives an estimated redshift of $\rm
z_{phot}=1.44$ for this source.  Besides, although it is not detected in the
V-band \footnote{even when its flux is manually measured with the IRAF task
`phot'}, it is a significant detection in the B-band ($\rm B=27.2 \pm 0.3$, AB).
This source  corresponds to object number 9 in the Stanway et al. list.

\item {\bf e1605}: this source has `hyperz' primary solution $\rm z_{phot}=6.71$
(with an associated probability $\rm P \approx 90\%$), but the BPZ code gives
$\rm z_{phot}=2.84$. This source is not detected on the z-band images, but it
could have some flux in the $\rm I_{775}$-band ($\rm I_{775}=27.9 \pm 1.3,
AB$). Its very red near-IR colour, i.e. $\rm (J-K_s)>2.5$ (Vega), suggests 
that the estimated redshift from the BPZ code is more reliable in this case.

\item{\bf e2006}: this source has two `hyperz' solutions with similar
probabilities. The primary solution is $\rm z_{phot}=6.21 \, (P \approx 73\%)$
and the secondary one is $\rm z_{phot}=1.15 \, (P \approx 60\%)$. The BPZ 
estimation is $\rm z_{phot}=6.37$. We measured $\rm (I_{775}-z)= 2.1 \pm 0.8 $
for this object and we do not find it is significantly detected in either the B
or V bands. However, the higher redshift solution would imply it is an extremely
bright galaxy with a k-corrected absolute magnitude $\rm M_{K_s}=-28.5$ and
estimated mass $\rm \sim 6.7 \times 10^{11} M_\odot$, so we consider that in
this case the `hyperz' secondary solution is more likely to be the right answer.

\end{itemize}

\begin{table*}
\caption{List of ERGs in the Roche et al. sample 
with `hyperz' primary solutions 
$\rm z_{phot} >4$. The first column is the identification number. The second and
third columns indicate the coordinates of the source obtained from the $\rm
K_s$-band images. The fourth and fifth columns show the redshifts estimated by
`hyperz' and the BPZ code, respectively. Column six lists comments on the individual objects.}
\begin{tabular}{rcccccl}
\hline \hline ERG id & RA(J2000) & DEC(J2000) & hyperz & BPZ & accepted? &
COMMENTS \\ \hline 114 & 3:32:22.46 & -27:50:47.16 & 5.60$\rm \small
^{+0.03}_{-0.14}$ & 5.75 & N & Unresolved. $(\rm J-K_s)_{Vega}=0.8 \pm 0.3$ \\
566 &3:32:24.78 & -27:49:12.91 & 6.69$\rm \small ^{+0.11}_{-0.12}$ & 1.28 & N &
Unresolved. $(\rm J-K_s)_{Vega}=1.1 \pm 0.3$. $\rm B_{AB}=23.7 \pm 0.1$ \\ 967 &
3:32:18.17 & -27:47:46.48 & 5.39$\rm \small ^{+0.03}_{-0.08}$ & 5.70 & N &
Spectroscopically confirmed star. $\rm V_{AB}=27.9 \pm 0.4$ \\ 1504 &3:32:18.17
& -27:46:16.33 & 4.63$\rm \small ^{+0.10}_{-0.06}$ & 1.44 & N & $\rm B_{AB}=27.2
\pm 0.3$ \\ 1605 &3:32:19.67 & -27:46:02.04 & 6.71$\rm \small ^{+0.75}_{-0.32}$
& 2.84 & N & $\rm (J-K_s)_{Vega}>2.5$. Not detected in z band.\\ 2386
&3:32:19.23 & -27:45:45.39 & 5.33$\rm \small ^{+0.05}_{-0.04}$ & 5.56 & N &
Unresolved. $(\rm J-K_s)_{Vega}=1.2 \pm 0.2$\\ 2006 & 3:32:28.81 & -27:44:30.54
& 6.21$\rm \small ^{+0.17}_{-0.33}$ & 6.37 & N & Significant secondary solution $\rm z_{phot}=1.15$ \\ 778 & 3:32:41.76 & -27:48:24.92 & 4.74$\rm
\small ^{+0.20}_{-0.25}$ & 5.42 & N & Revised UDF photometry implies  $\rm
z_{phot}=1.56$\\ 1113 & 3:32:34.65 & -27:47:20.89 & 5.07$\rm \small
^{+0.16}_{-0.08}$ & 1.05 & N & Revised UDF photometry implies  $\rm
z_{phot}=1.43$ \\ 1272 & 3:32:41.69 & -27:46:55.40 & 5.03$\rm \small ^{+0.21}_{-0.13}$ & 5.48 & N & Revised UDF photometry implies  $\rm
z_{phot}=1.91$ 
\\ 491 & 3:32:25.95 &
-27:49:30.38 & 4.58$\rm \small ^{+0.24}_{-0.09}$ & 5.22 & Y & No significant
`hyperz' secondary solutions.
\\ 1703 & 3:32:37:74 &
-27:45:05.41 & 4.74$\rm \small ^{+0.04}_{-0.05}$ & 5.38 & Y & No significant
`hyperz' secondary solutions.\\ \hline \hline
\end{tabular}
\end{table*}

\begin{figure}
\begin{center}
\includegraphics[width=1.0\hsize,angle=0] {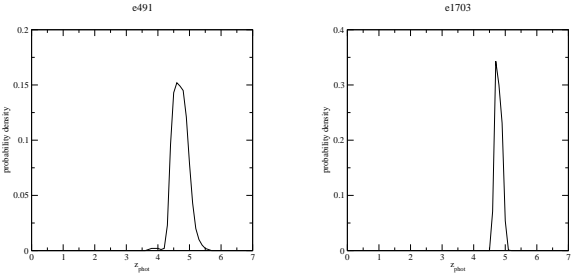}
\caption[]{ \label{fig-highzcomp} Redshift probability density 
distributions for the accepted $\rm z>4$ candidate sources 
in the Roche et al. ERG sample. }
\end{center}  
\end{figure}

\noindent 3/12 high-z candidate ERGs lie in the Hubble UDF, which images have been recently made publicly available. After revision of the photometry on the HST-ACS and HST-NICMOS ultra-deep optical and near-IR images, we reject also these 3/12  high-z candidates:

\begin{itemize}
\item {\bf e778}: using 5-epoch GOODS HST-ACS (V1.0) and VLT-ISAAC data, the `hyperz' primary solution for this source is $\rm z_{phot}=4.74$ (with an
associated probability $\rm P \approx 98\%$). The BPZ redshift estimation
is $\rm z_{phot}=5.42$. However, after revision of the photometry on the Hubble UDF images, this source appears to be a significant detection in the B-band. The revised  `hyperz' primary solution is $\rm z_{phot}=1.56$.

\item {\bf e1113}: this source initially had a `hyperz' primary solution $\rm z_{phot}=5.07$
($\rm P \approx 56\%$) and secondary solution $\rm z_{phot}=1.10$ ($\rm P
\approx 26\%$). The BPZ code favoured a low redshift solution $\rm
z_{phot}=1.05$. We did not detect this source in either the B or V bands on the GOODS V1.0 images. However, it is a clear detection in both the B and V bands on the Hubble UDF images. The revised `hyperz' primary solution is $\rm z_{phot}=1.43$. This source corresponds to object number 2 in Stanway et al. list.

\item{\bf e1272}: this source initially had a `hyperz' primary solution $\rm z_{phot}=5.03$
( $\rm P \approx 82\%$) and no significant secondary solution. The BPZ 
estimation was $\rm z_{phot}=5.48$.  It was not a significant detection in either the B or V bands on the GOODS V1.0 images, but it is on the Hubble UDF maps.
The revised `hyperz' primary solution is $\rm z_{phot}=1.91$.

\end{itemize} 

\noindent We accept the remaining 2/12 as  $\rm z>4$ candidates, whose redshift probability density distributions we show in fig. \ref{fig-highzcomp}:


\begin{itemize}
\item {\bf e491}: `hyperz' primary solution is $\rm z_{phot}=4.58$ (with an
associated probability $\rm P \approx 98\%$). The BPZ code estimates a redshift of 
$\rm z_{phot}=5.22$. There are no significant secondary solutions for the
redshift of this source in `hyperz'. The $\rm K_s$-band k-corrected absolute
magnitude for this galaxy is $\rm M_{K_s}=-25.6$ and we estimate a mass of $\rm 5.8
\times 10^{10} M_\odot$.  This object has $\rm (I_{775}-z)= 0.6 \pm 0.3$ and it
is not detected in either the B or V bands.

\item{\bf e1703}: this source has a `hyperz' primary solution $\rm z_{phot}=4.74$
( $\rm P \approx 90\%$) and no significant secondary solution. The BPZ 
redshift is $\rm z_{phot}=5.38$. The estimated $\rm K_s$-band k-corrected
absolute magnitude for this source is $\rm M_{K_s}=-27.3$ and the estimated mass
is $\rm 2.7 \times 10^{11} M_\odot$ . It has $\rm (I_{775}-z)= 0.97 \pm 0.14$
and it is not detected in either the B or V bands.

\end{itemize} 

     Finally, we list other ERGs in the Roche et al. sample which are among the
     very-high redshift candidates selected by other authors, and for which we
     obtain lower redshift estimates:

\vspace{0.5cm}

\begin{tabular}{rccl}
\hline ERG id & Author & Author's id & `hyperz'\\ \hline 82 & Bremer et al. & 7
& 1.40$\rm \small ^{+0.30}_{-0.04}$\\ 309 & Bremer et al. & 6 & 2.32$\rm \small
^{+0.11}_{-0.05}$\\ 225 & Dickinson et al. & SiD013 & 1.74$\rm \small
^{+0.08}_{-0.10}$ \\ 1423 & Bremer et al. & 2 & 1.60$\rm \small
^{+0.28}_{-0.08}$\\ \hline
\end{tabular}
\vspace{0.5cm}

\noindent From the sources listed above, only e309 has a very-high-redshift
secondary solution, although with negligible significance, at $\rm z_{phot}=6.74$.

   To summarise, we conclude that, within the limits of our survey,  only a few ERGs are likely to be at very high redshifts. Deeper near-IR selected samples of galaxies are necessary to confirm or refute the rarity of massive objects assembled at very early epochs.

\subsection{Review of the ERG cluster candidate at redshift $\rm z \sim 1.5$}

   Roche et al. (2003) found 10 ERGs within a $20^{\prime \prime}$ 
   radius of the Chandra source
   XID:58 (RA=3:32:11.85, DEC=-27:46:29.14, J2000), which is itself a counterpart of
   the ERG identified as e1435 in Roche et al. list. This overdensity
   suggested the presence of an ERG cluster at redshift $\rm z \sim 1.5$, based on the 
   photometric redshift of the Chandra source XID:58, estimated
   as $\rm z_{phot}=1.44$ by Mainieri (2003).  Based on the source
   colours, Roche et al. argued that 7/9 of the ERGs around XID:58 (e1435) could
   be part of that cluster.  In the following table, we present our redshift
   estimates (both primary and secondary) for e1435 and the 7 surrounding sources 
   that could be associated with it.

\vspace{0.5cm}
\begin{tabular}{lcc}
\hline ERG id & $\rm z_{phot}$ (primary) & $\rm z_{phot}$ (sec.) \\ \hline 1333
& 3.10$\rm \small ^{+0.11}_{-0.18}$ & 1.78\\ 1341 & 1.52$\rm \small
^{+0.06}_{-0.13}$ & 1.01 \\ 1390 & 1.29$\rm \small ^{+0.01}_{-0.01}$ & 1.67 \\
1404 & 1.52$\rm \small ^{+0.02}_{-0.03}$ & 1.78 \\ 1423 & 1.60$\rm \small
^{+0.28}_{-0.09}$ & 1.34 \\ 1435 (XID:58) & 1.85$\rm \small ^{+0.08}_{-0.08}$ &
1.30 \\ 1464 & 3.19$\rm \small ^{+0.07}_{-0.10}$ & 1.48 \\ 1481 & 1.30$\rm
\small ^{+0.12}_{-0.12}$ & 1.01 \\ \hline
\end{tabular}
\vspace{0.5cm}

\noindent Our estimated redshifts seem to confirm the presence of a cluster at
redshift $\rm z_{phot} \sim 1.5$. However, it is not clear whether all the
sources listed above belong to it. e1333 and e1464 seem to be higher redshift
sources, although the `hyperz' secondary solution might locate e1464 at redshift
$\rm z_{phot}=1.48$.  For e1435 (counterpart of XID:58) we estimate a redshift
$\rm z_{phot}=1.85$, a value somewhat higher than Mainieri's
estimation. Our secondary solution, on the contrary, is closer to their
value. The sources which seem to be part of a cluster at redshift $\rm z_{phot}
\sim 1.5$ are e1341, e1404, e1423 and e1481.

   Roche et al. also suggested that the 2/9 remaining ERGs around XID:58 could
   be associated with another X-ray source (Chandra XID:149), which is
   spectroscopically confirmed to be at redshift $\rm z_{phot}=1.033$ (Szokoly et al. 2003). Our estimated redshifts for these 2 remaining ERGs
   are:

\vspace{0.5cm}
\begin{tabular}{lcc}
\hline ERG id & $\rm z_{phot}$ (primary) & $\rm z_{phot}$ (sec.) \\ \hline 1311
& 1.05$\rm \small ^{+0.02}_{-0.02}$ & 0.61 \\ 1467 & 1.07$\rm \small
^{+0.04}_{-0.04}$ & 5.24 \\ \hline
\end{tabular}
\vspace{0.5cm}

\noindent The `hyperz' primary solutions seem to confirm that these 2 ERGs are
actually associated with the X-ray source XID:149.

\section{Summary and conclusions}

\parskip=0pt

In this paper we have presented estimated redshifts for the Extremely 
Red Galaxies selected by Roche et al. (2003) in the 50.4 arcmin$^2$ of the 
GOODS/CDFS deep ISAAC field to a limiting magnitude $\rm K_s=22$. 
This is the deepest significant sample of ERGs selected to date and 
constitutes a complement to other shallower but wider surveys of ERGs 
in the same field (Moustakas et al. 2004).

We have used multicolour photometry in seven passbands (B,V, $\rm I_{775}$, z, J, H and $\rm K_s$) to compute photometric redshifts using the public code 
`hyperz'. The B,V, $\rm I_{775}$ and z magnitudes have been measured on the 
stacked 5-epoch GOODS ACS images (V1.0 data release). We have obtained the 
redshift distribution and the corresponding Hubble diagram 
($\rm K_s-z_{phot}$), which show the existence of ERGs up to redshifts 
$\rm z_{phot} \sim 4.75$ at this depth. The ERG population is characterised 
by a large dispersion in the $\rm K_s-z_{phot}$ relation. We find that this 
dispersion is even more important in the dust-corrected Hubble diagram, 
indicating that the ERG populaton is composed of objects spanning a 
wide range in mass. We estimate that the ERGs in Roche et al.
sample span two decades in mass, from $\rm \sim 3 \times 10^9 M_\odot$ to $\rm
\sim 3 \times 10^{11} M_\odot$.  

We have studied the red envelope of galaxy evolution, determining the 
galaxy template which best describes the ERG dust-corrected 
reddest colours observed in the Roche et al. sample as a function of redshift.
We find that a starburst formed at redshift $\rm z_f=5$ with
passive evolution thereafter and solar metallicity provides a very 
good description of the red envelope of ERG evolution for objects 
selected with $\rm K_s<22$. 
Our work indicates that the simplified traditional
picture for ERGs as either old elliptical galaxies
or young dusty starbursts can no longer be sustained. The colours of many of
the reddest ERGs can only be explained by the combined effects of evolved
stellar populations and dust.

We have also explored the evolution of the ERG luminosity function from 
redshifts $\rm \langle z_{phot} \rangle =1$ to $\rm \langle z_{phot} 
\rangle =2.5$. The ERG LF at redshifts $\rm 1<z_{phot}<2$ reproduces the 
shape of the bright end of the global $\rm K_s$-band LF, as perhaps 
expected.  We find no evolution in the bright end of the LF of ERGs from 
redshifts $\rm \langle z_{phot} \rangle =1.0$ to $\rm \langle z_{phot} 
\rangle =2.5$ ($\rm -26<M_{K_s}<-25$) and from redshift $\rm \langle z_{phot} 
\rangle =2.0$ to $\rm \langle z_{phot} \rangle  =2.5$ ($\rm -27<M_{K_s}<-26$).
This effect cannot be deduced from inspection of the Hubble diagram for ERGs, 
constructed using the raw $\rm K_s$ magnitudes. Only after the application of 
dust corrections are some ERGs revealed as bright and massive galaxies.

One of the main goals in the selection of ERG samples is to trace the progenitors of 
very luminous local galaxies ($\rm L>L^\ast$). Only a small
fraction ($\rm \sim 15 \%$) of the ERGs in the Roche et al. 
sample can evolve into such luminous galaxies under passive evolution.
At redshifts $\rm z_{phot}  \sim 1-2$,  we find that the  comoving density of ERGs progenitors of massive galaxies is less than $\sim 1/10 $ of the local value. Between redshifts $\rm z=0$ and $\rm z=1.5$,  $\rm \Lambda CDM$ models only predict a decrease in the comoving densities of  dark matter halos massive enough as to host local $ \rm L>L^\ast$ galaxies by a factor $\sim$ 1.5-2.   This suggests that either ERGs cannot account for the whole population of local massive galaxies or the passive evolution scenario is not completely valid.  This deficiency appears as being more dramatic at lower ($\rm z_{phot} \sim 1-2$) than higher ($\rm z_{phot} \sim 3-4$) redshifts, suggesting that ERGs could be rather better tracing the progenitors of local massive galaxies present at early epochs.

Finally, we have investigated the existence of Lyman break galaxies 
massive enough to be included in an ERG sample. Out of an initial candidate list of 12 sources, we only confirm 2 ERGs as having  high 
probabilities of being located at redshifts $\rm z_{phot}>4$. 
Deep observations in wider areas are crucial in order to
 constrain lower limits on the fraction of massive galaxies with the
bulk of their stellar mass already assembled at very high redshifts.

Several open questions still remain for the ERG population in the 
GOODS/CDFS field.  An important issue is to understand how ERG morphology 
changes with redshift, a problem we plan to study in a future paper. Also, 
very deep spectroscopic observations scheduled with the Gemini Multiobject 
Spectrograph (GMOS) will provide more accurate redshifts for a subset of the 
ERGs in the Roche et al. sample, allowing a detailed study of their properties.


\section*{Acknowledgements}
This paper is based on observations made with the Advanced Camera for Surveys  and the Near Infrared Camera and Multi Object Spectrometer on
board the Hubble Space Telescope operated by NASA/ESA and with the Infrared
Spectrometer and Array Camera on the `Antu' Very Large Telescope operated by the
European Southern Observatory in Cerro Paranal, Chile, and form part of the
publicly available GOODS datasets.  We thank the GOODS teams for providing
reduced data products. We thank Will Percival for providing us a code to compute comoving densities of dark matter halos.  We also thank the referee, Tomonori Totani, for his suggestions and comments, which enriched the discussion of results in this paper.  
 
 KIC aknowledges funding from a POE-network studentship and the Overseas
 Research Scheme Award (ORS/2001014037). JSD, RJM and NDR acknowledge PPARC
 funding.


\bibliographystyle{mn2e}

\section*{References}

\bib Aretxaga I., Hughes D. H., Chapin E. L., Gazta\~naga E.,
 Dunlop J. S., Ivison R. J., 2003, MNRAS, 342, 759

\bib Ben\'{\i}tez N., 2000, ApJ, 536, 571

\bib Bertin E., Arnouts S., 1996, A\&A, 117, 393

\bib Bolzonella M., Miralles J.-M., Pell\'o R., 2000, A\&A, 363, 476

\bib Bolzonella M., Pell\'o R., Maccagni D., 2002, A\&A, 395, 443

\bib Bouwens R. J., Illingworth G. D., Rosati P., Lidman C. et al., 2003,
ApJ, 595, 589 (astro-ph/0306215)

\bib Bremer M.N., Lenhert M.D., Waddington I., Hardcastle M.J., Boyce P.J.,
Phillipps S., 2004, MNRAS, 346, L7 (astro-ph/0306587)

\bib Bruzual A. G., Charlot S., 1993, ApJ, 405, 538

\bib Bruzual G., Charlot S., 2003, MNRAS, 344, 1000

\bib Calzetti D., Armus L., Bohlin R. C., Kinney A. L., Koornneef
J., Storchi-Bergmann T., 2000, ApJ, 533, 682

\bib Chapman S. C., Blain A. W., Ivison R. J., Smail I. R., 2003, Nature, 
422, 695

\bib Charlot S., Fall S. M., 2000, ApJ, 539, 718

\bib Cimatti A., Andreani P., R\"{o}ttgering H., Tilanus R., 1998, Nature, 392,
895

\bib Cimatti A., Daddi E., Mignoli M., Pozzetti L. et al., 2002, A\&A, 381, L68

\bib Cole S., Norberg P., Baugh C. M., Frenk C. S. et al., 2001,
MNRAS, 326, 255

\bib Dey A., Graham J. R., Ivison R. J., Smail I., Wright G. S.,
Liu M. C., 1999, ApJ, 519, 610

\bib Dickinson M., Giavalisco M., The GOODS Team, 2003, in: The Mass of
Galaxies at Low and High Redshift, ESO Astrophysics Symposia, p. 324,
eds. Bender R. \& Renzini A., Springer.

\bib Dickinson M., Stern D., Giavalisco M., Ferguson H. C. et al., 2004, ApJ, 600, L99 (astro-ph/0309070)

\bib Dunlop J., Peacock J., Spinrad H., Dey A., Jim\'{e}nez R., Stern D.,
Windhorst R., 1996, Nature, 381, 581

\bib Dunlop J.S., 2001, in van Bemmel I.M., Wilkes B., Barthel P., ed.,
 FIRSED2000, New Astronomy Reviews, 45, Elsevier, p.609

\bib Elston R., Rieke G.H., Rieke M.J., 1988, ApJ, 331,77

\bib Frayer D.T., Reddy N. A., Armus L., Blain A.W., Scoville N. Z., Smail I.,
2004, AJ, 127, 728  (astro-ph/0310656)

\bib Kauffmann G., Charlot S., 1998, MNRAS, 297, L23

\bib Kochanek C.S. et al., 2001, ApJ, 560, 566

\bib Maihara T., Iwamuro F., Tanabe H., Taguchi T. et
al., 2001, PASJ, 53, 25

\bib Mainieri V., 2003, PhD thesis, Univ. Roma Tre

\bib Marzke R.O., da Costa L.N., Pellegrini P.S., Willmer C.N.A., Geller M.J., 1998, ApJ, 503, 617

\bib Miyazaki M. et al., 2003, PASJ, 55, 1079

\bib Moustakas L., Casertano S., Conselice C., Dickinson M.
E. et al., 2004, ApJ, 600, L131 (astro-ph/0309187)

\bib Roche N. D., Dunlop J., Almaini O., 2003, MNRAS, 346, 803
(astro-ph/0303206)

\bib Saracco P., Giallongo E., Cristiani S., D'Odorico S., Fontana A., Iovino
A., Poli F., Vanzella E., 2001, A\&A, 375, 1

\bib Saracco P., Longhetti M., Severgnini P., Della Ceca R., Mannucci F., Bender
R., Drory N., Feulner G., Ghinassi F., Hopp U., Maraston C., 2003, A\&A, 398,
127

\bib Saracco P. et al. 2004, A\&A, 420, 125 (astro-ph/0310131)

\bib Smail I., Owen F. N., Morrison G. E., Keel W. C., Ivison R. J., Ledlow 
M. J., 2002, ApJ, 581, 844

\bib Soifer B.T., Matthews K., Neugebauer G., Armus L., Cohen J.G., Persson
S.E., Smail I., 1999, AJ, 118, 2065

\bib Somerville R. S., Primack J. R., 1999, MNRAS, 310, 1087

\bib Somerville R. S., Lee K., Ferguson H. C., Gardner J. P., Moustakas L. A., Giavalisco M., 2004, ApJ, 600, L171 (astro-ph/0309071)

\bib Stanway E. R., Bunker A. J., McMahon R. G., 2003, MNRAS, 342, 439 (astro-ph/0302212)

\bib Szokoly G. P., Bergeron J., Hasinger G., Lehmann I. et al., 2003, submitted to ApJ supplement (astro-ph/0312324)

\bib Totani T., Yoshii Y., Iwamuro F., Maihara T.,
Motohara K., 2001, ApJ, 558, L87

\bib Wehner E. H., Barger A. J., Kneib J.-P., 2002, ApJ, 577, L83

\bib Willott C. J., Rawlings S., Jarvis M. J., Blundell K. M.,
2003, MNRAS, 339, 173

\bib Yan L., Thompson D., 2003, ApJ, 586, 765

 \end{document}